\documentclass{article}
\usepackage{amsmath}

\input{tcilatex}
\begin{document}

\title{The quantum speed up as advanced knowledge of the solution}
\author{Giuseppe Castagnoli \\
Pieve Ligure (Genova)\ giuseppe.castagnoli@gmail.com}
\maketitle

\begin{abstract}
With reference to a search in a database of size $N$, Grover states: \textit{%
"What is the reason that one would expect that a quantum mechanical scheme
could accomplish the search in }$O\left( \sqrt{N}\right) $\textit{\ steps?
It would be insightful to have a simple two line argument for this without
having to describe the details of the search algorithm"}. The answer
provided in this work is: "\textit{because any quantum algorithm takes the
time taken by a classical algorithm that knows in advance 50\% of the
information that specifies the solution of the problem". }In database
search, knowing in advance 50\% of the $n$\ bits that specify the database
location, brings the search from $O\left( 2^{n}=N\right) $ to $O\left(
2^{n/2}=\sqrt{N}\right) $. This empirical rule, unnoticed so far, holds for
both quadratic and exponential speed ups and is theoretically justified in
three steps: (i) once the physical representation is extended to the
production of the problem on the part of the oracle and to the final
measurement of the computer register, quantum computation is \textit{%
reduction} on the solution of the problem under a relation representing
problem-solution interdependence, (ii) the speed up is explained by a simple
consideration of time symmetry, it is the gain of information about the
solution due to backdating, to before running the algorithm, a
time-symmetric part of the reduction on the solution; this advanced
knowledge of the solution reduces the size of the solution space to be
explored by the algorithm, (iii) if $\mathcal{\Im }$ is the information
acquired by measuring the content of the computer register at the end of the
algorithm, the quantum algorithm takes the time taken by a classical
algorithm that knows in advance 50\% of $\mathcal{\Im }$, which brings us to
the initial statement.

The fact that a problem solving and computation process can be represented
as a single interaction, sheds light on our capability of perceiving
(processing) many things together at the same time in the so called
"present".
\end{abstract}

\section{Premise}

We represent an entire problem solving and computation process as a single
interaction: the reduction of the forward evolution (ending with the state
before measurement) on the backward evolution (the same unitary
transformation but ending with the state after measurement) under a relation
representing problem-solution interdependence\footnote{%
We do not require that reduction is objective in character. What is
considered to be objective here is the whole quantum process -- preparation,
unitary evolution, and measurement -- in fact represented as a single
interaction. We use the term "reduction" in its mathematical meaning of
projection on the subspace of an eigenvalue of the measured observable.}.
The relation applies to an infinite set of variables representing the
amplitudes of the computational basis vectors throughout the quantum
process; reduction changes these variables from the values assumed in the
forward evolution to the values assumed in the backward evolution. This
representation emphasizes the interdependence between all the parts of the
information processing and turns out to be useful in two areas: the speed up
of the quantum algorithms (their higher efficiency with respect to the
corresponding classical algorithms) and the unity of perception.

For what concerns the quantum algorithms, we extend their physical
representation to the production of the problem on the part of the oracle
and to the measurement of the computer register at the end of the algorithm.
The quantum process becomes reduction on the solution of the problem under
the problem-solution interdependence relation. This highlights an empirical
fact unnoticed so far: a quantum algorithm takes the time taken by a
classical algorithm that knows in advance 50\% of the information acquired
in the final measurement of the computer register. The reader interested in
this rule can go to section 4. For its theoretical explanation, he should
also read section 3.

The fact that an entire problem solving and computation process can be
represented as a single interaction, sheds light on the unity of perception.
The reader interested in this part can go directly to section 5, which also
provides the links to the notions developed in the previous sections. Here
below we provide an overview of the work.

\section{Overview}

Solving a problem requires two steps. A problem solving step (deriving, from
the formulation of the problem, the solution algorithm) and a computation
step (running the algorithm). The latter step is generally oblivious of the
former. We unify the two steps into a single physical interaction,
represented as a hypothetical many body interaction in the classical
framework (a useful conceptual reference), as a measurement interaction in
the quantum framework, as follows.

For the conventional representation of computation, we adopt the so called
"circuit model". At the logical level, a set of Boolean variables
(representing the logical state of the computer register) undergoes a
sequence of elementary reversible/deterministic input output
transformations, changing input values into output values. Through this
sequence, an overall input is transformed into an overall output
representing the solution of the problem. The sequence can be represented as
a \textit{linear Boolean network}, a series of reversible Boolean gates with
no feedback loops and no preassigned values on the outputs, where the input
logically propagates to the output in a deterministic way. At the physical
level, a computer register undergoes a sequence of deterministic
transformations (corresponding to the logical ones) that change input states
of the register into output states until the overall output is reached.
Reading the register content in the overall output state yields the solution
of the problem. This representation of computation is oblivious of the
problem (in the quantum algorithms, of the oracle that produces the
problem), an omission that makes the speed up conceptually unexplained.

Computation is unified with problem solving as follows. We consider the%
\textit{\ non-linear Boolean network} representing the original problem, to
be distinguished from the above said linear network, which represents the
algorithm that solves the problem. The nonlinear network is generally
exponentially smaller, has logically irreversible gates or partial gates,
feedback loops (outputs feeding back into inputs), and outputs with
preassigned values. Solving this network means finding a Boolean assignment
that satisfies all network elements (gates, wires, preassigned Boolean
values). The non linear network is not directly solvable by the
deterministic propagation of an input into an output; it is usually
transformed into an exponentially larger linear network (representing the
algorithm that solves the problem) solvable by this kind of propagation.
Here we proceed in a different way, identifying a physical interaction that
directly produces the solutions of the nonlinear network. We replace the
Boolean variables $x_{i}$ of the nonlinear network by ratios between real
non-negative variables $X_{i}/Q$, the network constraints by equations on
these ratios. For $Q>0$, the solutions of this system of equations
correspond to the solutions of the nonlinear network. The relation between
variables (the ratios) established by the system of equations represents
problem-solution interdependence. The solutions are produced by a quantum
measurement interaction, first visualized as a hypothetical classical many
body interaction\footnote{%
This visualization allows to bring to quantum computation the machanical
engineering notion that the absence of dissipation of a process does not
imply its mechanical invertibility.}. This latter is inspired to a well
known paradox of classical mechanics: statically, the application of
external forces to a perfectly rigid body is balanced by infinitely many
distributions of stress inside the body, against one distribution if the
body is flexible. This paradox is ported to a perfectly rigid body made of
moving parts, whose coordinates $X_{i}$ and $Q$ are submitted to mechanical
constraints representing the system of equations. In the initial
configuration all coordinates are zero. By applying a force to the "input"
part of coordinate $Q$, the many distributions of stress inside the body
find a combination of movements of the machine parts that satisfies all the
constraints at the same time. In more detail, pushing the input part from $%
Q=0$ to $Q>0$, brings in the many body problem, the non-determination of the
dynamics (because of perfect coincidence of interaction times between many
bodies). By applying the principle of sufficient reason, we postulate a many
body interaction that produces a solution of the problem with probability
proportional to the mass of the machine parts moved to produce that
solution. In reality, the kinematics and the statistics of this hypothetical
classical interaction represent a quantum measurement interaction.
Configuration space becomes phase space. The ratios $X_{i}/Q$ represent the
populations of (the reduced density operators of) the qubits of a quantum
register in the surrounding of measurement. The many body interaction is
replaced by the measurement interaction, which changes these population
variables from the values immediately before to the values immediately after
measurement. The problem-solution interdependence relation linearly extends
to an infinite set of variables representing the amplitudes of the
computational basis vectors throughout the quantum process (these latter
variables are function of the above population variables). Under the
extended relation, the measurement interaction changes the forward
evolution, ending with the state before measurement, into the backward
evolution, ending with the state after measurement that encodes the solution
of the problem. In this "relational" representation, quantum computation is 
\textit{reduction} on the solution under the relation representing
problem-solution interdependence.

Applied to the quantum algorithms, this representation of problem solving
and computation (for short, computation from now on) explains the quantum
speed up, the fact that some quantum algorithms are much faster than the
corresponding classical algorithms; the physical representation of
computation must be extended to the production of the problem on the part of
the oracle and to the final measurement of the register's content. Then any
known quantum algorithm becomes reduction on the solution under the relation
representing problem-solution interdependence. The quantum speed up is
explained by a simple consideration of time-symmetry. It is the gain of
information about the solution of the problem due to backdating, to before
running the algorithm, a time-symmetric part of the reduction on the
solution. This advanced knowledge of the solution\footnote{%
We have chosen the term "advanced" because it is reminiscent of the
"advanced wave", going backward in time, of Cramer's transactional
interpretation of quantum mechanics.} reduces the size of the solution space
to be explored by the algorithm: let $\mathcal{\Im }$\ be the information
acquired\ by measuring the content of the computer register at the end of
the algorithm, the quantum algorithm takes the time taken by a classical
algorithm that knows in advance 50\% of $\mathcal{\Im }$. This is verified
for the algorithms of Deutsch, Grover, and Simon, and also holds in the case
of the generalized Simon's algorithm, thus for the hidden subgroup
algorithms.

The notion that an entire problem solving and computation process can be
condensed into a single interaction (an idealized classical many body
interaction or a quantum measurement interaction) can also explain the unity
of perception, the fact that we perceive (assumedly, process) many things
together at the same time in the so called "present". Tacking into account
many constraints at the same time is exactly what a classical many body
interaction, or a quantum measurement interaction, does. The physical
representation of the notion of "present" is the instant of the interaction
in the classical case, the time interval spanned by backdated reduction in
the quantum case.

\section{The notion of relational computation}

We develop the representation of computation as reduction of the forward
evolution on the backward evolution under a relation representing
problem-solution interdependence.

\subsection{The notion of fundamental relation in classical and quantum
physics}

The idea that many things are processed all together at the same time,
standing at the basis of relational computation, is formalized by resorting
to a notion of the Gestalt theory (e. g. Mulligan and Smith, 1988). The
wholeness/unity of a physical situation implies that there is a relation --
a "simultaneous dependence" in the language of the theory -- between all the
quantitative variables describing it.

An example of fundamental relation in classical physics is "force equal mass
times acceleration" in the case of a point mass. In view of what will
follow, it should be noted that this relation is implicitly assumed to be
objectively perfect. If we see it as a mechanism, whose degrees of freedom
are the variables related by the law, this mechanism should be perfectly
accurate, rigid, and reversible -- it is not the case that Newton's second
law gets deformed and jams because of mechanical flexibility or
irregularity, or dissipates because of friction.

Another important feature of the fundamental relations that we find in
Nature, is that they can be nonfunctional, which is also the case of
Newton's law. The change of any one variable is correlated with an identical
change of the product or ratio of the other two variables but does not
determine their individual changes. Correspondingly, Newton's law can host
nondeterminism in the form of the many body problem.

An example of functional simultaneous dependence in quantum physics is the
correlation between the polarizations of two photons in an entangled
polarization state. We should note an important difference with respect to
the classical notion of simultaneity. Now, if the two measurements are
successive in time, there is simultaneous dependence between two results
displaced in time (this is of course the notion of correlation between
measurement outcomes in an entangled state). Simultaneous dependence is time
symmetric: it is not the case that the former result determines the latter,
causality between the two results is mutual.

An example of nonfunctional simultaneous dependence is the partial OR
relation between \ the three orthogonal components of a spin $1$\ particle.

The requirement that the relation representing problem-solution
interdependence is perfect, is essential in the case of the classical many
body interaction, it is absorbed into the quantum principle and the notion
of qubit (Finkelstein, Coral Gables Conference, 1969 and Phys. Rev., 1969)
in the case of the measurement interaction. That infinite classical
precision can be dispensed for because of quantization has already been
noted by Finkelstein (Finkelstein, 2008).

\subsection{Relational computation as a hypothetical classical many body
interaction}

We postulate a many body interaction inspired to a well known paradox of
classical mechanics: statically, the application of external forces to a
perfectly rigid body is balanced by infinitely many distributions of stress
inside the body, against one distribution if the body is flexible. This
paradox is ported to a perfectly rigid body made of moving parts, whose
coordinates are submitted to mechanical constraints representing the
problem. Applying a force to an "input part" brings in the many body
problem. It is reasonable to postulate that the many distributions of stress
inside the body find a combination of movements of the body's parts that
satisfies all the constraints at the same time.

It is interesting to note that giving up the limitation to two body
interaction marks the departure from classical computation. The fundamental
physical model of classical computation is the bouncing ball model of
reversible computation (Fredkin and Toffoli, 1982). Here the variables at
stake are ball positions and momenta. Outside collisions, there is no
simultaneous dependence between the variables of different balls, which are
independent of each other. During collision, there is simultaneous
dependence between the variables of the colliding balls, but this is
confined to ball pairs (there can be several collisions at the same time,
but involving independent ball pairs, with no simultaneous dependence
between the variables of different pairs). The simultaneous collision
between many balls is avoided to avoid the many body problem, the
non-determination of the dynamics.

Instead, by assuming a perfect simultaneous dependence between all
computational variables, one can devise an idealized classical machine that,
thanks to a many body interaction, nondeterministically produces the
solution of a (either linear or non linear) system of Boolean equations
under the simultaneous influence of all equations.

Let us start with the simple problem of finding the solutions ($x=0$, $y=1$
and $x=1$,$~y=0$) of the single Boolean equation $y=\overline{x}$. The
mechanical principle that produces these solutions can easily be extended to
any system of Boolean equations. Let $Q$,$~X$, and$~Y$ be real non negative
variables. The Boolean problem is transformed into the problem of finding
the solutions, for$~Q>0$, of the system of equations%
\begin{equation}
Q=X+Y,  \label{linear}
\end{equation}%
\begin{equation}
Q^{2}=X^{2}+Y^{2}.  \label{quadratic}
\end{equation}

$Q=0$ implies$~$ $X=Y=0$, while $\frac{X}{Q}$ and $\frac{Y}{Q}$ are
undetermined. When $Q>0$, $\frac{X}{Q}$ coincides with the Boolean variable $%
x$ and$~\frac{Y}{Q}$ with $y=\overline{x}$. Equations (\ref{linear}) and (%
\ref{quadratic}), representing the problem constraint $y=\overline{x}$,
establish a nonfunctional relation, a simultaneous dependence, between the
variables $Q$, $X$, and $Y$. To introduce the notion of relational
computation, it is useful to think that the solutions are produced under
this relation through a many body interaction as follows. We put a
differential gear between coordinate $Q$ (the input of the gear) and
coordinates $X$ and $Y$ (the two outputs of the gear), which implements
equation (\ref{linear}). We put another differential gear between the
squares of these coordinates, namely between the auxiliary coordinates $%
Q^{\prime }$, $X^{\prime }$, and $Y^{\prime }$ connected through parabolic
cams to $Q$, $X$, and $Y$, so that $Q^{\prime }=Q^{2}$, $X^{\prime }=X^{2}$,
and $Y^{\prime }=Y^{2}$, which implements equation (\ref{quadratic}).

The initial machine configuration is $Q=X=Y=0$; it can be argued that any
motion of the part of coordinate $Q$ from $Q=0~$to$~Q>0$ produces a solution
in a nondeterministic way, as follows. The many body problem is the problem
of the non-determination of the dynamics in the case of perfect coincidence
between interaction times of many bodies -- which is the case if we try and
push part $Q$ out of $Q=0$. Here we postulate a solution to the many body
problem by applying the principle of sufficient reason. The motion of part $%
Q $ could be obtained by applying a force to it. In fact, there is no reason
for either $X$ or $Y$, in a mutually exclusive way, not to move with $Q$,
since either movement offers zero static resistance to the force -- there is
only the inertia of the machine parts.

By playing with inertia, we can also tune the probabilities that $X$ and $Y$
move. For example, we can ask that, on average (over an ensemble of
repetitions of the interaction), there is equipartition of energy among the
machine degrees of freedom. This implies that the probability that the parts
(either $Q$\ and $X$\ or $Q$ and$\ Y$) move, is proportional to the mass of
the parts. Under this assumption, and by assuming for example even masses
for $X$\ and $Y$, the values of the coordinate ratios change from$\ \frac{X}{%
Q}=\frac{Y}{Q}=\frac{1}{2}$ before interaction to $\frac{X}{Q}=1$ and $\frac{%
Y}{Q}=0$ or (in a mutually exclusive way) $\frac{X}{Q}=0$ and $\frac{Y}{Q}=1$
after interaction.

This hypothetical behavior is an arbitrary prolongation of classical
computation to the many body case. In reality, it describes quantum
measurement, as shown in the next section.

We should note that, unlike deterministic reversible processes, the present
process is not invertible. For example, we can think of connecting the input
part to an ideal spring charged when $Q=0$. On the one side, there would be
oscillations without dissipation. On the other, at each oscillation, the
movement of the input part from $Q=0~$to$~Q>0$ would randomly drag either $%
X~ $or$~Y~$in a mutually exclusive way. \ 

The extension of this computation mechanism to the problem of solving a
generic system of Boolean equations is provided in the Appendix.

We should note that this form of computation is essentially different from
the causal propagation of an input into an output. For example, it can
produce two inputs such that their product is a preassigned output (the
nonlinear Boolean network for this problem is the network for the
multiplication of two integer numbers, with a preassigned value on the
output). If this were an input-output propagation (which is not), one should
say that inputs are produced with advanced knowledge of the output.

\subsection{Relational computation as a quantum process}

The kinematics and the statistics of the classical many body interaction can
describe a quantum measurement interaction. Considering the previous example 
$y=\overline{x}$, the motion from $Q=0$ to $Q>0$ is analogous to measuring
two qubits $X$ and $Y$ in the entangled state $\left\vert 0\right\rangle
_{X}\left\vert 1\right\rangle _{Y}+e^{i\delta }\left\vert 1\right\rangle
_{X}\left\vert 0\right\rangle _{Y}$, where $\delta $\ is an arbitrary (even
random) phase. The configuration space of the classical machine becomes the
phase space of the two qubits, the normalized coordinates of the machine
parts (the ratios $\frac{X}{Q}$, $\frac{Y}{Q}$) represent the populations of
(the reduced density operators of) the qubits at time $t_{r}$, at the end of
the unitary evolution stage of the quantum process -- we are thus dealing
with "population variables". The measurement interaction changes these
population variables from the values assumed in the foward evolution to the
values assumed in the backward evolution. The correspondence between
normalized coordinates and population variables is:

\begin{equation}
\frac{X}{Q}=x_{00}\left( t_{r}\right) =1-x_{11}\left( t_{r}\right) ,~\frac{Y%
}{Q}=y_{00}\left( t_{r}\right) =1-y_{11}\left( t_{r}\right) ,
\label{correspondence}
\end{equation}%
where $x_{00}\left( t_{r}\right) $ is the variable representing the
population of qubit $X$ in $\left\vert 0\right\rangle _{X}\left\langle
0\right\vert _{X}$ during the interaction, etc. The problem-solution
interdependence relation is the same as before: 
\begin{equation}
Q=X+Y,  \label{qlinear}
\end{equation}%
\begin{equation}
Q^{2}=X^{2}+Y^{2}.  \label{qquadratic}
\end{equation}%
In the transition from $Q=0~$to $Q>0$, the population variables change from
the values before reduction $x_{00}\left( t_{r}\right) =$ $x_{11}\left(
t_{r}\right) =$ $y_{00}\left( t_{r}\right) =$ $y_{11}\left( t_{r}\right) =$ $%
\frac{1}{2}$ to the values after reduction, one of the two mutually
exclusive sets of values $x_{00}\left( t_{r}\right) =0,~x_{11}\left(
t_{r}\right) =1,$ $y_{00}\left( t_{r}\right) =1,~y_{11}\left( t_{r}\right) =$
$0$ and $x_{00}\left( t_{r}\right) =1,~x_{11}\left( t_{r}\right) =0,$ $%
y_{00}\left( t_{r}\right) =0,~y_{11}\left( t_{r}\right) =$ $1$.

It should be noted that simultaneous dependence (functionally) extends to
the amplitudes of the basis vectors throughout the unitary evolution stage
of the quantum process -- for example the process of entangling $X$\ and $Y$
starting from a sharp preparation of the two qubits. Amplitudes should be
represented by variables, function of the above population variables. To
this end, we define the state of the quantum system at time $t_{r}$ as a
function of the population variables:

\begin{equation}
\left\vert \psi ,t_{r}\right\rangle =\sum_{i=1}^{n}\alpha _{i}\left(
t_{r}\right) \left\vert i\right\rangle ,  \label{evolution}
\end{equation}%
where $\left\vert i\right\rangle $\ is the $i$-th computational basis vector
and

\begin{equation}
\alpha _{i}\left( t_{r}\right) =f_{i}\left[ x_{00}\left( t_{r}\right)
,~y_{00}\left( t_{r}\right) \right] .  \label{function}
\end{equation}%
The transition from $Q=0~$to $Q>0$ changes together the population variables 
$x_{00}\left( t_{r}\right) ,~y_{00}\left( t_{r}\right) $ and the amplitudes $%
\alpha _{i}\left( t_{r}\right) $ from the values assumed in the forward
evolution to the values assumed in the backward evolution. The entire
unitary evolution can be represented as a unitary transformation of $%
\left\vert \psi ,t_{r}\right\rangle $:

\begin{equation}
\forall t:\left\vert \psi ,t\right\rangle =U(t,t_{r})^{\dag }\left\vert \psi
,t_{r}\right\rangle ,  \label{unitary}
\end{equation}%
where $U(t,t_{r})$ is the unitary transformation undergone by the quantum
system from $t$\ to $t_{r}$. Under the relation established by equations (%
\ref{correspondence}), (\ref{qlinear}), (\ref{qquadratic}), (\ref{evolution}%
), (\ref{function}), and (\ref{unitary}), reduction changes the forward
evolution into the backward evolution. This way of representing together the
unitary evolution and the measurement stage of the quantum process -- as the
reduction of the forward evolution onto the backward evolution under a
relation representing problem-solution interdependence, is essential for the
explanation of the speed up.

It should be noted that, under the above relation,\ any two amplitudes $%
\alpha _{i}\left( t_{1}\right) $ and $\alpha _{j}\left( t_{2}\right) $, with 
$t_{1}\leq t_{2}$, also depend from one another in a time symmetric way. In
other words, it is not the case that the change (from the forward to the
backward value) of the amplitude at time $t_{1}$ causes the change of the
amplitude at the later time $t_{2}$; causality is mutual, like in the
measurement of two entangled polarizations. In this sense, we can say that
there is simultaneous dependence between all the amplitudes of the
computational vectors throughout the entire quantum process.

\section{Relational computation and the quantum algorithms}

We apply relational computation to the quantum algorithms. It suffices to
represent together (as a single physical interaction) the production of the
problem on the part of the oracle, the unitary evolution stage of the
quantum process, and the final measurement of the content of the computer
register. This shows that quantum computation is reduction on the solution
of the problem under the problem-solution interdependence relation, and
explains the speed up.

We need to consider only the input and the output of the unitary
transformation performed by the quantum algorithm, therefore no previous
knowledge of the subject is required to the reader. For an introductory
description of the quantum algorithms in the same notation used here, see
(Kaye et al., 2007).

\subsection{Grover's algorithm}

The problem addressed by Grover's algorithm (Grover 1996) is database
search. It can be seen as a game between two players with a chest of $N$\
drawers; the first player (the oracle) hides a ball in drawer number $k$.
The second player must find where the ball is. Opening drawer $x$ to check
whether the ball is in it amounts to computing the Kronecker function $%
\delta \left( k,x\right) $, which is $1$ if $k=x$ and $0$ otherwise. Here $%
k\in \left\{ 0,1\right\} ^{n}$, where $n=\log _{2}N$ (for simplicity, we can
assume that $N$ is a power of $2$).

The value of $k$ chosen by the first player is hardwired inside a black box
that, for each input $x$, computes $\delta \left( k,x\right) $. This black
box is given to the second player, who has to find the value of $k$\ by
computing $\delta \left( k,x\right) $ for different values of $x$ (i. e., by
opening different drawers to check whether the ball is in it). In the
classical case, to find the value of $k$, $\delta \left( k,x\right) $\ must
be computed the order of $N$ times, in the quantum case the order of $\sqrt{N%
}$ times -- there is a quadratic speed up

We review the original Grover's algorithm. The first player chooses a black
box with a given value of $k$, and gives it to the second player. Instead of
trying, one by one, single values of $x$, the second player prepares an $n$%
-qubit register $X$ in an even weighted superposition of all the possible
values of $x$, and computes $\delta \left( k,x\right) $ in quantum
parallelism. For example, with $N=4$ and $k=01$, the algorithm unitarily
changes the input state

\begin{equation}
\frac{1}{2\sqrt{2}}\left( \left\vert 00\right\rangle _{X}+\left\vert
01\right\rangle _{X}+\left\vert 10\right\rangle _{X}+\left\vert
11\right\rangle _{X}\right) \left( \left\vert 0\right\rangle _{V}-\left\vert
1\right\rangle _{V}\right) ,  \label{in}
\end{equation}%
into the output state

\begin{equation}
\frac{1}{\sqrt{2}}\left\vert 01\right\rangle _{X}\left( \left\vert
0\right\rangle _{V}-\left\vert 1\right\rangle _{V}\right)  \label{out}
\end{equation}%
(the quantum network implementing this transformation, as well as the
function of register $V$, can be disregarded here). Measuring the content of
register $X$\ in the output state (\ref{out})\ yields the solution. In the
present case ($N=4$), this is obtained by computing $\delta \left(
k,x\right) $ only once.

Thus, in the original algorithm, the solution is obtained in a deterministic
way (for simplicity, we put ourselves in those values of $N$\ where the
probability of error of Grover's algorithm is zero); relational computation,
namely reduction on the solution of the problem under the problem-solution
interdependence relation, is completely hidden. This is because the random
generation of $k$ is not represented physically.

We extend the physical representation by adding an ancillary $n$-qubit\
register $K~$prepared in an even weighted superposition of all the possible
values of $k$. The black box that, given an input $x$, computed $\delta
\left( k,x\right) $ for a specific value of $k$, is now replaced by a black
box that, given the inputs $k$\ and $x$, computes $\delta \left( k,x\right) $%
. The extended algorithm computes $\delta \left( k,x\right) $ as before, but
now for a superposition of combinations of values of $k$ and $x$. The input
state is now:

\begin{equation}
\text{ }\frac{1}{4\sqrt{2}}\left( \left\vert 00\right\rangle _{K}+\left\vert
01\right\rangle _{K}+\left\vert 10\right\rangle _{K}+\left\vert
11\right\rangle _{K}\right) \left( \left\vert 00\right\rangle
_{X}+\left\vert 01\right\rangle _{X}+\left\vert 10\right\rangle
_{X}+\left\vert 11\right\rangle _{X}\right) (\left\vert 0\right\rangle
_{V}-\left\vert 1\right\rangle _{V}),  \label{preparation}
\end{equation}%
where the superposition hosted in register $K$ can indifferently be coherent
or incoherent (in which case each element of the superposition should be
multiplied by a random phase factor). The extended algorithm unitarily
transforms the input state (\ref{preparation})\ into the output state:%
\begin{equation}
\frac{1}{2\sqrt{2}}\left( \left\vert 00\right\rangle _{K}\left\vert
00\right\rangle _{X}+\left\vert 01\right\rangle _{K}\left\vert
01\right\rangle _{X}+\left\vert 10\right\rangle _{K}\left\vert
10\right\rangle _{X}+\left\vert 11\right\rangle _{K}\left\vert
11\right\rangle _{X}\right) (\left\vert 0\right\rangle _{V}-\left\vert
1\right\rangle _{V}),  \label{final}
\end{equation}%
where each value of $k$ is entangled with the corresponding solution found
by the second player (the same value of $k$ but in register $X$). The final
measurement of the contents of registers $K$\ and $X$ in state (\ref{final}%
)\ determines the moves of both players. The reduction induced by measuring
the content of register $K$, backdated to before running the algorithm,
yields the original Grover's algorithm. Thus, by representing together the
production of the problem on the part of the oracle, the unitary evolution,
and the measurement of the register content, we find reduction on the
solution of the problem under the problem-solution interdependence relation.
The present relation is a straightforward generalization (from two one-qubit
registers to two two-qubit registers) of the relation established by
equations (\ref{correspondence}), (\ref{qlinear}), (\ref{qquadratic}), (\ref%
{evolution}), (\ref{function}), and (\ref{unitary}) -- section 3.3.

The nondeterministic production of the contents of the two registers by
quantum measurement, can be seen as mutual determination between such
contents, like in the measurement of two entangled polarizations. This
justifies the square root speed up, as follows. We cannot say that reading
the content of $K$ at the end of the algorithm determines the content of $X$%
, namely that choosing the drawer to hide the ball in determines the drawer
the ball is found in -- this would be the classical game with no mutual
determination. For the same reason we cannot say that reading the content of 
$X$ determines the content of $K$, that looking inside a drawer at the end
of the algorithm creates the ball in it. Mutual determination is
symmetrical, it should be represented by saying that the contents of the two
registers are determined by reading the first (second) bit of register $K~$%
and the second (first) bit of register $X$.

Then Grover's algorithm is equivalent to the following game. We arrange the $%
N$ drawers in a matrix of $\sqrt{N}$\ columns and $\sqrt{N}$ rows. At the
end of the algorithm, the first player determines (say) the row by reading
the first bit of register $K$. The second player determines the column by
reading the second bit of register $X$, say that this reading is $1$. The
reduction induced by the second player, backdated to before running the
algorithm, changes the initial preparation of register $K$, $\frac{1}{2}%
\left( \left\vert 00\right\rangle _{K}+\left\vert 01\right\rangle
_{K}+\left\vert 10\right\rangle _{K}+\left\vert 11\right\rangle _{K}\right) $
(eq. \ref{preparation}), into $\frac{1}{\sqrt{2}}\left( \left\vert
01\right\rangle _{K}+\left\vert 11\right\rangle _{K}\right) $, thus
determining the column before running the algorithm. In this picture,
Grover's algorithm searches just the row randomly chosen by the first
player, which justifies the $\sqrt{N}$ computations of $\delta \left(
k,x\right) $ -- the quadratic speed up. Grover's algorithm is equivalent to
a classical search in a database of size $\sqrt{N}$ (we should symmetrize
for the exchange of columns and rows). See also (Castagnoli, 1997),
(Castagnoli et al., 1999), (Castagnoli and Finkelstein, 2001).

The same justification holds in the case that the value of $k$ is already
determined before running the algorithm, like in virtual database search.
This case is indistinguishable from the random generation of $k$ at the end
of the algorithm, where backdating reduction makes $k$ predetermined.

If we think that $k$ is predetermined, the initial superposition in register 
$K$ -- $\frac{1}{2}\left( \left\vert 00\right\rangle _{K}+\left\vert
01\right\rangle _{K}+\left\vert 10\right\rangle _{K}+\left\vert
11\right\rangle _{K}\right) $ -- represents the initial ignorance of the
value of $k$ on the part of the second player. Its reduction to $\frac{1}{%
\sqrt{2}}\left( \left\vert 01\right\rangle _{K}+\left\vert 11\right\rangle
_{K}\right) $, due to the second player reading $1$ in the second bit of
register $X$ at the end of the algorithm, gives -- before running the
algorithm -- an information gain of one bit\footnote{%
If we assume that the initial superposition hosted in register $K$ is
incoherent, this information gain is the decrease of the von Neumann entropy
of the (reduced density operator of the) content of register $K$ during
measurement.}. This is 50\% of the information acquired by reading the
solution, namely the two bits of register either $K$\ or $X$ (the
information content of one register is redundant with respect to the content
of other register), or one bit of $K$\ and the other bit of $X$.

For a database of size $N$, the reduction of ignorance about the solution
due to backdating, to before running the algorithm, 50\% of the information
acquired by reading the solution, is:%
\begin{equation}
\Delta _{\mathcal{\Im }}=\frac{1}{2}\lg _{2}N.  \label{first}
\end{equation}

Summing up, in the relational representation:

\begin{enumerate}
\item \bigskip Quantum computation is \textit{reduction} on the solution of
the problem under the relation representing problem-solution interdependence.

\item The speed up is the reduction of the initial ignorance of the solution
due to backdating, to before running the algorithm, a time-symmetric part of
the reduction on the solution, namely 50\% of the information acquired by
reading the solution. This advanced knowledge of the solution reduces the
size of the solution space to be explored by the algorithm.

\item Let $\mathcal{\Im }$ be the information acquired by measuring the
content of the computer register at the end of the algorithm; the quantum
algorithm takes the time taken by a classical algorithm (the\textit{\
reference classical algorithm}) that knows in advance 50\% of $\mathcal{\Im }
$.
\end{enumerate}

In the present case of Grover's algorithm, knowing in advance 50\% of the $n$
bits that specify the database location, reduces the size of the solution
space (of the database) from $2^{n}=N$ to $2^{n/2}=\sqrt{N}$. The quantum
algorithm working on a database of size $N$\ takes the time taken by a
classical algorithm working on a database of size $\sqrt{N}$. See also
(Castagnoli, 2007, 2008, 2008).

We should note that the information acquired by measuring the content of the
computer register at the end of the algorithm, specifies both the problem
(the oracle's random choice) and the solution. Therefore, "reduction on the
solution of the problem" should be understood as reduction on the solution
and the problem. However, in Grover's algorithm, the information acquired by
reading the solution and the problem coincides with the information acquired
by reading either one, since either one is a function of the other.

We should also note that point \textbf{3}, besides being theoretically
justified within the relational representation of computation, is an
empirical fact holding for all quantum algorithms -- once the physical
representation is extended to the production of the problem on the part of
the oracle and to the final measurement of the register's content.

\subsection{Deutsch's algorithm}

We consider the revisitation of Deutsch algorithm (Deutsch, 1985) due to
Cleve et al.(1996). Now the oracle chooses at random one of the four binary
functions $f_{\mathbf{k}}:\left\{ 0,1\right\} \rightarrow \left\{
0,1\right\} $ (see table \ref{table}\ -- $\mathbf{k}\equiv k_{1}k_{2}$ is a
two-bit string belonging to $\left\{ 0,1\right\} ^{2}$):

\begin{equation}
\begin{tabular}{|c|c|c|c|c|}
\hline
$x$ & $f_{00}(x)$ & $f_{01}(x)$ & $f_{10}(x)$ & $f_{11}(x)$ \\ \hline
0 & 0 & 0 & 1 & 1 \\ \hline
1 & 0 & 1 & 0 & 1 \\ \hline
\end{tabular}%
.  \label{table}
\end{equation}%
Note that $k_{1}=f_{\mathbf{k}}(0)$ and $k_{2}=f_{\mathbf{k}}(1)$. Then the
oracle gives to the second player the black box hardwired for the
computation of that function. The second player, by trying function
evaluation for different values of $x$, must find out whether the function
is balanced (i. e. $f_{01}$ or $f_{10}$, with an even number of zeroes and
ones) or constant (i. e. $f_{00}$ or $f_{11}$). This requires two function
evaluations in the classical case, just one in the quantum case (this has
been the first speed up ever found).

In the conventional quantum algorithm, the second player prepares two
one-qubit registers $X$\ and $V$ in the input state

\begin{equation}
\frac{1}{2}\left( \left\vert 0\right\rangle _{X}+\left\vert 1\right\rangle
_{X}\right) \left( \left\vert 0\right\rangle _{V}-\left\vert 1\right\rangle
_{V}\right)
\end{equation}%
(the function of register $V$\ can be disregarded here). With just one
function evaluation, the algorithm unitarily produces the output state

\begin{equation}
\frac{1}{\sqrt{2}}\left\vert 0\right\rangle _{X}\left( \left\vert
0\right\rangle _{V}-\left\vert 1\right\rangle _{V}\right)
\end{equation}%
if the function is constant and

\begin{equation}
\frac{1}{\sqrt{2}}\left\vert 1\right\rangle _{X}\left( \left\vert
0\right\rangle _{V}-\left\vert 1\right\rangle _{V}\right)
\end{equation}%
if the function is balanced. Thus, at the end of the algorithm, register $X$
contains the solution of the problem ($0$ $\equiv $ constant, $1$ $\equiv $\
balanced). The speed up is obtained in a deterministic way, but also in this
case the random generation of $\mathbf{k}$ is not represented physically. We
extend the physical representation by adding an ancillary two-qubit\
register $K~$prepared in a superposition (indifferently coherent or
incoherent) of all the possible values of $\mathbf{k}$. The input state is
now:%
\begin{equation}
\frac{1}{4}\left( \left\vert 00\right\rangle _{K}+\left\vert 01\right\rangle
_{K}+\left\vert 10\right\rangle _{K}+\left\vert 11\right\rangle _{K}\right)
\left( \left\vert 0\right\rangle _{X}+\left\vert 1\right\rangle _{X}\right)
\left( \left\vert 0\right\rangle _{V}-\left\vert 1\right\rangle _{V}\right) .
\end{equation}

The extended algorithm, given the inputs $\mathbf{k}$\ and $x$, computes $f(%
\mathbf{k},x)\equiv f_{\mathbf{k}}(x)$, yielding the output:

\begin{equation}
\frac{1}{2\sqrt{2}}\left[ \left( \left\vert 00\right\rangle _{K}-\left\vert
11\right\rangle _{K})\left\vert 0\right\rangle _{X}+(\left\vert
01\right\rangle _{K}-\left\vert 10\right\rangle _{K})\left\vert
1\right\rangle _{X}\right) \right] \left( \left\vert 0\right\rangle
_{V}-\left\vert 1\right\rangle _{V}\right) .  \label{fin}
\end{equation}%
The measurement of the content of registers $K$\ and $X$\ in the output
state determines the moves of both players (the random choice of the
function on the part of the first player and the answer provided by the
second player). The information acquired is $2$ bits -- the two bits of
register $K$ (the content of register $X$\ is a function of the content of $%
K $, therefore the information contained in $X$ is redundant). The quantum
algorithm takes the time taken by a classical algorithm working on a
solution space reduced in size because one bit of information about the
content of register $K$, either $k_{1}=f_{\mathbf{k}}(0)$ or $k_{2}=f_{%
\mathbf{k}}(1)$, is known in advance. This algorithm must acquire the other
bit of information by computing either $f_{\mathbf{k}}(1)$ or $f_{\mathbf{k}%
}(0)$. Thus the reference classical algorithm has to perform just one
function evaluation like the quantum algorithm. This verifies point \textbf{3%
}.

\subsection{Simon's and the hidden subgroup algorithms}

In Simon's algorithm, a first player (the oracle) chooses at random a
function among the set of the "periodic" functions $f_{\mathbf{k}}:\left\{
0,1\right\} ^{n}\rightarrow S$, where $S\subseteq \left\{ 0,1\right\} ^{n}$.
The "periodic" function $f_{\mathbf{k}}$, where $\mathbf{k=~}k_{1}...k_{n}$
is a string of $n$\ Boolean values (excluding the all zeroes string), is
such that $f_{\mathbf{k}}\left( x\right) =f_{\mathbf{k}}\left( y\right) $ if
and only if $x=y$\ or $x=y\mathbf{\oplus k}$. Here $x$ and $y$ are variables
belonging to $\left\{ 0,1\right\} ^{n}$\ (also represented as $n$\ bit
strings) and $\oplus $\ denotes bitwise addition modulo $2$ (see the
following example). Then he gives to the second player a black box that,
given an input $x$, computes $f_{\mathbf{k}}\left( x\right) $. The second
player should find the hidden string $\mathbf{k}$\ (among $2^{n}-1$\
possible strings)\ through function evaluation. To find the value of $%
\mathbf{k}$ with probability, say, $\frac{2}{3}$, $f_{\mathbf{k}}\left(
x\right) $ must be computed the order of $2^{\frac{n}{3}}$ times in the
classical case, the order of $3n$ times in the quantum case. There is an
exponential speed up.

Let us exemplify. With $n=2$ and $S=\left\{ 0,1\right\} $, there are three
"periodic" functions (up to permutation of function values leaving the
hidden string unaltered):%
\begin{equation}
\begin{tabular}{|c|c|c|c|}
\hline
$x$ & $f_{01}\left( x\right) $ & $f_{10}\left( x\right) $ & $f_{11}\left(
x\right) $ \\ \hline
00 & 0 & 0 & 0 \\ \hline
01 & 0 & 1 & 1 \\ \hline
10 & 1 & 0 & 1 \\ \hline
11 & 1 & 1 & 0 \\ \hline
\end{tabular}%
.  \label{tables}
\end{equation}

The original Simon's algorithm is as follows. The second player prepares an
\ $n$-qubit register $X$ in an even weighted superposition of all the
possible values of $x$, and an \ $m$-qubit register $F$\ (devoted to contain
the result of function evaluation) in a sharp state (standing for a blank
register). In the present example, the input state of the algorithm is thus:

\begin{equation}
\frac{1}{2}\left( \left\vert 00\right\rangle _{X}+\left\vert 01\right\rangle
_{X}+\left\vert 10\right\rangle _{X}+\ \left\vert 11\right\rangle
_{X}\right) \left\vert 0\right\rangle _{F}.  \label{initial}
\end{equation}%
Then he performs function evaluation on the superposition of all the
possible values of $x$, obtaining the intermediate output (say that $\mathbf{%
k}=10$, see $f_{10}\left( x\right) $\ in table \ref{tables}):

\begin{equation}
\frac{1}{2}\left[ \left( \left\vert 00\right\rangle _{X}+\left\vert
10\right\rangle _{X}\right) \left\vert 0\right\rangle _{F}+\left( \left\vert
01\right\rangle _{X}+\left\vert 11\right\rangle _{X}\right) \right]
\left\vert 1\right\rangle _{F}.  \label{evaluation}
\end{equation}%
Now he applies the Hadamard transform (still a unitary transformation) to
the state of register $X$, obtaining the overall output:%
\begin{equation}
\frac{1}{2}\left[ \left( \left\vert 00\right\rangle _{X}+\left\vert
01\right\rangle _{X}\right) \left\vert 0\right\rangle _{F}+\left( \left\vert
00\right\rangle _{X}-\left\vert 01\right\rangle _{X}\right) \right]
\left\vert 1\right\rangle _{F}.  \label{hadamard}
\end{equation}%
In the overall output state (\ref{hadamard}), for each value of the
function, register $X$ hosts an even weighted superposition of the $2^{n-1}$
strings $\mathbf{h}_{j}=h_{j1}h_{j2}...h_{jn}$ orthogonal to $\mathbf{k}$ --
such that, for all $j$, $\left( \sum_{i=1}^{n}h_{ji}k_{ji}\right) $ modulo $%
2=0$; in the example, $\mathbf{h}_{1}\equiv 00$\ and $\mathbf{h}_{2}\equiv
01 $\ \ are the two strings orthogonal to $\mathbf{k\equiv }$ $10$. Note
that, in (\ref{hadamard}), only the phase of the even weighted amplitudes
depend on the value of $f_{\mathbf{k}}\left( x\right) $. Therefore, by
measuring the content of $X$ in (\ref{hadamard}), one obtains at random one
of the $\mathbf{h}_{j}$. The entire process (initial preparation of
registers $X$\ and $F$, unitary transformation, and measurement of the
content of $X$) is iterated until obtaining $n-1$ different $\mathbf{h}_{j}$%
, which allows to find $\mathbf{k}$ by solving a system of $n-1$ modulo $2$
linear equations\footnote{%
Throughout this article, the term "solution" should be unerstood in the
broader sense of "information acquired at the end of the quantum algorithm",
here leading to the solution of the problem with a classical post processing.%
}.

This formulation of Simon's algorithm leaves the number of iterations of the
algorithm unbounded. Alternatively, we can iterate the algorithm a fixed
number of times, which leaves a certain probability of failing to find the
solution. If the algorithm is iterated, say, $6n$ times, the probability of
obtaining $n-1$ different $\mathbf{h}_{j}$, thus of finding the solution, is
about $\frac{8}{9}$ (e. g. Kaye et al., 2007). The computation time taken by
the $6n$ iterations Simon's algorithm, in terms of number of elementary
logical operations, is the order of $n^{3}$, against the order of $2^{n}$ of
the classical algorithm.

Now we extend Simon's algorithm to represent the random choice of the hidden
string on the part of the oracle. We add an auxiliary $n$-qubit register $K$%
, prepared in an even weighted superposition of the $2^{n}-1$ possible
values of $\mathbf{k}$. The black box that (given input $x$) computed $f_{%
\mathbf{k}}\left( x\right) $ for a specific $\mathbf{k}$, is now replaced by
a black box that, given the inputs $\mathbf{k}$ and $x$, computes $f\left( 
\mathbf{k},x\right) \equiv f_{\mathbf{k}}\left( x\right) $. The input state
is now:

\begin{equation}
\frac{1}{2\sqrt{3}}\left( \left\vert 01\right\rangle _{K}+\left\vert
10\right\rangle _{K}+\ \left\vert 11\right\rangle _{K}\right) \left(
\left\vert 00\right\rangle _{X}+\left\vert 01\right\rangle _{X}+\left\vert
10\right\rangle _{X}+\ \left\vert 11\right\rangle _{X}\right) \left\vert
0\right\rangle _{F},  \label{einitial}
\end{equation}%
where the superposition hosted in register $K$ is indifferently coherent or
incoherent. The overall output state (\ref{hadamard}) becomes now:\newline
\begin{eqnarray}
&&+\frac{1}{2}\left\vert 01\right\rangle _{K}\left[ \left( \left\vert
00\right\rangle _{X}+\left\vert 10\right\rangle _{X}\right) \left\vert
0\right\rangle _{F}+\left( \left\vert 00\right\rangle _{X}-\left\vert
10\right\rangle _{X}\right) \left\vert 1\right\rangle _{F}\right]  \notag \\
&&+\frac{1}{2}\left\vert 10\right\rangle _{K}\left[ \left( \left\vert
00\right\rangle _{X}+\left\vert 01\right\rangle _{X}\right) \left\vert
0\right\rangle _{F}+\left( \left\vert 00\right\rangle _{X}-\left\vert
01\right\rangle _{X}\right) \left\vert 1\right\rangle _{F}\right]
\label{extended} \\
&&+\frac{1}{2}\left\vert 11\right\rangle _{K}\left[ \left( \left\vert
00\right\rangle _{X}+\left\vert 11\right\rangle _{X}\right) \left\vert
0\right\rangle _{F}+\left( \left\vert 00\right\rangle _{X}-\left\vert
11\right\rangle _{X}\right) \left\vert 1\right\rangle _{F}\right] .  \notag
\end{eqnarray}%
\newline
The random choice of a value of $\mathbf{k}$\ on the part of the first
player is obtained by measuring the content of register $K$ in (\ref%
{extended}). This puts register $K$ in a sharp state and registers $X$\ and $%
F$ in a state of the form (\ref{hadamard}). The second player measures the
content of register $X$ \ (thus collecting the first of the $\mathbf{h}_{j}$%
) then, leaving register $K$ in its sharp state and working only on
registers $X$\ and $F$, iterates up to a total of $6n$ times the original
Simon's algorithm (thus collecting the other $6n-1$ values of $\mathbf{h}%
_{j} $).

In the case that the algorithm finds the solution (when there are at least $%
n-1$\ different $\mathbf{h}_{j}$, the probability of this being $\frac{8}{9}$%
), there is mutual determination between the content of register $K$\ and
the content of register $X$\ in each of its $6n$ measurements. In more
detail: measuring the content of $K$ projects the state of register $K$\ on
a single value of $\mathbf{k}$; which value of $\mathbf{k}$ is the result of
mutual causality between measuring the content of $K$ and the successive
measurements of the content of $X$. This is completely similar to measuring
the polarizations of two photons in an entangled polarization state; the
polarization measured first is the result of mutual causality between this
measurement and the successive measurement of the second polarization; it is
not the case that the first result determines the second, nor that the
second determines the first, causality is mutual even across an interval of
time.

Now, backdating 50\% of the information acquired by reading the content of
register $X$ (50\% of $6n$\ readings), reduces the size of the problem not
with certainty (like in the case of Grover's and Deutsch's algorithms) but
with probability $\frac{2}{3}$ (with good approximation, the probability of
finding the hidden string with $3n$ readings is in fact $\frac{2}{3}=1-\sqrt{%
1-\frac{8}{9}}$). Moreover, with this probability, the size of the problem
is reduced from $2^{n}-1$ to $1$.

The fact that the reference classical algorithm disposes of $n-1$\ different 
$\mathbf{h}_{j}$, implies that the quantum algorithm (whose $\mathbf{h}_{j}$
comprise those of the reference algorithm) also disposes of them: $\frac{2}{3%
}$ is thus the probability that both algorithms have enough information to
identify the solution. Point \textbf{3, }the fact that the quantum algorithm
takes the time taken by a reference classical algorithm that knows in
advance 50\% of the information about the solution\textbf{,} applies in a
probabilistic way. With probability $\frac{2}{3}$, we put ourselves in the
case that both algorithms dispose of $n-1$\ different $\mathbf{h}_{j}$ (the $%
\frac{1}{3}$ probability that this is not the case, goes down exponentially
fast with the number of iterations of the quantum algorithm, it is $\approx
\left( \frac{1}{3}\right) ^{g}$\ with $6gn$\ iterations). In this case, the
reference algorithm has simply to sort out $n-1$ different $\mathbf{h}_{j}$
and solve the related system of linear equations. Thus, the quantum
algorithm takes a time the order of $n^{3}$, the reference algorithm the
order of $n^{2}$, not so different in comparison with the order of $2^{n}$
taken by classical computation\footnote{%
The two errors, the lower probability of finding the solution and the lower
computational cost of the reference algorithm (with respect to the quantum
algorithm), tend to compensate with one another. The compensation is full in
the following perspective. We iterate the quantum algorithm other $6n$
times, just to feed the reference algorithm with other $3n$ values of $%
\mathbf{h}_{j}$. In this way the probability of finding the solution is the
same for the original $6n$\ iterations quantum algorithm and this enriched
reference algorithm. If we ascribe the cost of the enrichment entirely to
the reference algorithm, also computational costs are the same.}.

Perhaps a more significant way of comparing classical computation and the
reference algorithm is to say that:

\begin{enumerate}
\item[4.] classical computation has to find the hidden string by solving a
system of nonlinear equations in $n$\ variables (we can think of the
formulation of the problem in terms of real nonnegative variables), the
reference classical algorithm -- thanks to the advanced knowledge of 50\% of
the contents of register $X$ (measured in the successive iterations of the
algorithm) -- a system of linear equations in the same order of variables.
\end{enumerate}

The fact that points \textbf{1}, \textbf{2}, \textbf{3}, and \textbf{4} hold
for Simon's algorithm implies that they hold for a larger class of
algorithms, because of the tight similarity between Simon's algorithm, the
generalized Simon's algorithm, and the hidden subgroup formulation of the
quantum algorithms (Mosca and Ekert, 1999) -- like finding orders, finding
the period of a function (the quantum part of Shor's factorization
algorithm), finding discrete logarithms, etc. (Kaye, 2007). In all cases,
utilizing 50\% of the information acquired in a batch of iterations of the
quantum algorithm reduces the probability of finding the solution of an
amount that goes exponentially to zero with the number of iterations in the
batch, and reduces the problem of solving a nonlinear system in $n$\
variables to the problem of solving a linear system in a number of variables
of the same order. Point \textbf{4} holds replacing the expression "hidden
string " by "hidden subgroup".

\section{Relational computation and the unity of perception}

For wholeness of perception, as it appears in introspective analysis, I mean
the following. For example, in this moment, I see the room in which I am
working, an armchair, the window, the garden, and the Mediterranean Sea on
the background. In my visual perception, besides some aspects that are
addressed by artificial intelligence, like the recognition of patterns,
there is another thing that should be addressed by a physical information
theory, the both obvious and striking fact that I see so many things
together at the same time. What I see is close to a digital picture whose
specification would require a significant amount of information. And,
apparently, we can perceive a significant amount of information
simultaneously all together, in the so called "present". Another example is
our capability of grasping the solution of a problem. Reasonably, when we
grasp the solution, we should take into account at the same time the
statement of the problem, the solution, and the logical connection in
between.

In the assumption that perception is information processing, perceiving many
things at the same time poses the question: what form of computation can
process many things together at the same time in the so called "present",
and what is the physical counterpart of the introspective notion of
"present", such that an entire computation process can occur in it? The
mechanism of relational computation can provide an answer.

In the idealized classical case, an entire computation process is condensed
into an instantaneous many body interaction (sections 3.1, 3.2). The
physical counterpart of the introspective notion of "present" is here the
instant of the interaction.

In the quantum case, a state can hold any amount of information, which is
processed together "at the same time" by the sequence: preparation, unitary
transformation, and measurement. "At the same time" since there is
simultaneous dependence between all the amplitudes of the computational
basis vectors at any pair of times along the process (the change of one
amplitude, from the forward to the backward value, changes the other and
vice-versa, in a time symmetric way). Correspondingly the measurement
interaction changes the entire forward evolution into the backward evolution
(sections 3.1, 3.2, and 3.3). The physical counterpart of the notion of
"present" is here the time interval spanned by backdated reduction.

The observation that, in visual perception, we take into account many things
at the same time acquires a literal meaning. Taking into account many things
at the same time is exactly what many body interaction, or reduction of the
forward evolution on the backward evolution, does.

By the way, if the physical basis of consciousness is a fundamental
nondeterministic problem solving mechanism, it is difficult to think that
consciousness is the passive witness of a deterministic classical process.
Moreover, the deterministic, two-body character of classical computation
prevents taking into account many (so to speak, more than two) things
together at the same time, or (reasonably) hosting consciousness either.

The present model might provide some theoretical ground to the empirical
notion of premonition. Point \ \textbf{3} of sections 4.1 through 4.3, put
in anthropomorphic language, says that the quantum algorithm is quicker than
the classical algorithm because it disposes of an anticipated partial
knowledge of the solution it will find in the future -- knowing in advance
50\% of it. For the possibility that consciousness interacts with systems
displaced in time up to 500 milliseconds in the past and from milliseconds
to months into the future, see (Sheehan, 2008).

The present identification between the notions of simultaneous dependence,
physical law, and perception has a precedent in Plato's notion of Form (the
Greek word Eidos translates into Form, Idea, or Vision). In Phaedo: "Ideas
are objective perfections that exist in themselves and for themselves, at
the same time they are the cause of natural phenomena, they keep phenomena
bound together and constitute their unity" (Abbagnano, 1958). The Ideas of
our mind are clearly identified with physical laws; as well known, Platonic
Ideas are also perfect mathematical objects. The usual Platonist
interpretation of this ambivalence is that the mind can access an autonomous
and objective world of perfect mathematical ideas. A more physical
interpretation is the other way around, the ideas in our head -- our
perceptions -- are instances of physical laws, namely of objectively
perfect, nonfunctional simultaneous dependences.

The present idea that "grasping the solution of a problem" implies reduction
under a simultaneous dependence representing problem-solution
interdependence, is parallel to another statement of the theory of Forms:
"To know the Form of X is to understand the nature of X; so the philosopher
who, for example, grasps the Form of justice, knows not merely what acts are
just, but also why they are just" (Flew, 1984).

Also Plato's notion that the mind can access the objective perfection of the
world of ideas, while material objects are imperfect (a formulation of the
mind-body problem) is reflected in the present model, where perception is
the instantiation, in the quantum level of the brain, of an objectively
perfect physical law.

The present fundamental problem-solving mechanism matches in my judgement
with: (i) the Orchestrated Objective Reduction theory of consciousness of
Hameroff and Penrose (Hameroff and Penrose, 1996): it is a way of seeing
problem-solving, specifically of a form relying on the non linearity of
reduction, in Objective Reduction, (ii) the idea, also present in that
theory, that the very existence of consciousness depends on our capability
of accessing the Platonic world of objectively perfect mathematical Ideas
(Penrose, 1994), and (iii) Stapp's argument (Stapp, 1993) that consciousness
is incompatible with classical locality (here seen as two body interaction,
unable to process many things together at the same time) and compatible with
quantum non-locality and reduction (here seen as many body interaction, or
reduction under problem-solution interdependence, capable of processing many
things together at the same time).

\section{Conclusions}

In the admission of their same authors, the quantum speed ups are still
little understood. In his 2001 paper (Grover, 2001), Grover states: \textit{%
"What is the reason that one would expect that a quantum mechanical scheme
could accomplish the search in }$O\left( \sqrt{N}\right) $\textit{\ steps?
It would be insightful to have a simple two line argument for this without
having to describe the details of the search algorithm"}.

Casting the present argument in two lines, the answer is:

\textit{Because any quantum algorithm takes the time taken by a classical
algorithm that knows beforehand 50\% of the information specifying the
solution of the problem.}

The theoretical justification of this empirical fact can be segmented as
follows:

\begin{enumerate}
\item once the physical representation is extended to the production of the
problem on the part of the oracle, quantum computation is \textit{reduction}
on the solution of the problem under the relation representing
problem-solution interdependence;

\item the speed up is the reduction of the initial ignorance of the solution
due to backdating, to before running the algorithm, a time-symmetric part of
the reduction on the solution; this advanced knowledge of the solution
reduces the size of the solution space to be explored by the algorithm;

\item thus, if $\mathcal{\Im }$\ is the information acquired by measuring
the register content at the end of the algorithm, the quantum algorithm
takes the time taken by a classical algorithm that knows in advance 50\% of $%
\mathcal{\Im }$.
\end{enumerate}

The expounded mechanism could be used, by reverse engineering, for the
search of new quantum algorithms. For example, Grover's algorithm can be
seen as the symmetrization (for exchange of all the possible ways of getting
in advance the 50\% of $\mathcal{\Im }$) of a classical algorithm that does
database search in a solution space of quadratically reduced size. The 50\%
rule could be used to investigate which problems are liable of being solved
with a quadratic or exponential speed up, \ the question of why the speed up
is quadratic for unstructured problems, \ exponential for a very limited
class of structured problems, not others, etc. The relevance of the rule to
this question can be seen as follows. Let us think of a crossword puzzle,
the 50\% rule means knowing in advance 50\% of the characters of the word;
if the word is a random sequence of characters, this advanced knowledge
yields a quadratic speed up, if the word is structured, it can yield much
more.

From another standpoint, relational computation provides a comparate vision
of disparate forms of computation. One goes from deterministic classical
computation to nondeterministic quantum computation by giving up the
limitation to two body interaction.

It is interesting to compare reversible computation (Finkelstein, Coral
Gables Conference, 1969, Bennett, 1973 and 1982, Benioff, 1982, Fredkin and
Toffoli, 1982) with relational computation. In the present context, the
former form of computation can be seen as a special case of the latter:

\begin{itemize}
\item in reversible computation, the physical process is limited to the
unitary evolution and is thus deterministic in character; the two notions of
absence of dissipation and process invertibility overlap;

\item in relational computation, the physical process is extended to the
sequence unitary evolution and measurement and is thus nondeterministic in
character; the overlapping between the two notions is released, the absence
of dissipation can go together with the non-invertibility of the process
(section 3.2).
\end{itemize}

Summing up, nondeterminism (the non-invertibility of the process) here
becomes a fundamental feature of quantum computation. It is not loosing
track of a time-invertible propagation spreading through too many degrees of
freedom; it is capability of making choices satisfying many (like in many
body) constraints at the same time.

Furthermore, being the condensation of an entire computation process into a
single physical interaction (a classical many body interaction or a quantum
reduction), the present nondeterministic form of computation can represent
the information processing standing at the basis of perception. It explains
an essential feature of conscious perception, the fact that (e. g. in visual
perception) we see many things together at the same time. Taking into
account many constraints at the same time is exactly what classical many
body interaction, or quantum reduction of the forward evolution on the
backward evolution, does; "at the same time" is the instant of the
interaction in the classical case, the time interval spanned by backdated
reduction in the quantum case.

{\LARGE Appendix}

The computation mechanism expounded in section 3.2 is easily extended to
solve any system of Boolean equations, namely (without loss of generality) a
network of $n$ partial OR gates $\func{POR}(x_{i,1}~,x_{i,2},~x_{i,3})=1$ ($%
i=1,~...,~n$) and $m$\ wires $x_{i,j}=x_{h,k}$ (for $m$ assignments of $%
i,~j,~h,~k$). The truth table of the partial OR gate $i$ is given in the
right side of table (\ref{truth}).%
\begin{equation}
\begin{tabular}{|c|c|c|c|}
\hline
part & $x_{i,1}$ & $x_{i,2}$ & $x_{i,3}$ \\ \hline
$X_{i,1}$ & 0 & 1 & 1 \\ \hline
$X_{i,2}$ & 1 & 0 & 1 \\ \hline
$X_{i,3}$ & 1 & 1 & 0 \\ \hline
\end{tabular}
\label{truth}
\end{equation}%
The problem solving machine for this network is defined as follows. For all $%
i$, each row $j$ ($j=1,~2,~3$) of the truth table is associated with a
mechanical part of coordinate $X_{i,j}$ (table \ref{truth}) -- we also say
that part $X_{i,j}$ is labelled by the values of the Boolean variables $%
x_{i,1}~,x_{i,2},~x_{i,3}$ appearing in the corresponding row. $Q$ and $%
Q^{2} $\ (with a parabolic cam in between) are now the inputs of $n$\ pairs
of differential gears, one linear and the other nonlinear, as before. This
time, each linear gear $i$ has three outputs of coordinates $X_{i,j}~$($%
j=1,~2,~3$), each nonlinear gears their squares (with parabolic cams in
between), so that, for all $i$: $Q=\sum_{j=1}^{3}X_{i,j}$ and $%
Q^{2}=\sum_{j=1}^{3}X_{i,j}^{2}$. Therefore, the motions of the parts of
each \textit{triplet} ($X_{i,1}$, $X_{i,2}$, $X_{i,3}$) are mutually
exclusive with one another. If part $X_{i,j}$ moves with $Q$, we understand
that $x_{i,1}~,x_{i,2},~x_{i,3}$ assume the values appearing in the
corresponding row.

This is justified by the following implementation of the wires. For example,
let us assume that $x_{i,1}=x_{h,2}$, which means either $x_{i,1}=x_{h,2}=0$
or $x_{i,1}=x_{h,2}=1$. Looking at table (\ref{truth}), one can see that
this wire must be represented by the equations $X_{i,1}=X_{h,2}$ and $%
X_{i,2}+X_{i,3}=X_{h,1}+X_{h,3}$. In fact, if the part that moves in the
first triplet is $X_{i,1}$, this implies $x_{i,1}=0$ (see the intersection
between first row and first column of table \ref{truth}). Then, to satisfy
the wire, the part that moves in the second triplet must be $X_{h,2}$, so
that also $x_{h,2}=0$ (intersection between second row and second column of
table \ref{truth} -- having replaced the subfix $i$\ by $h$). This justifies
the first equation. If instead the part that moves in the first triplet is
either $X_{i,2}$ or $X_{i,3}$, this implies $x_{i,1}=1$ (intersection
between second or third row and first column). Then, to satisfy the wire,
the part that moves in the second triplet must be either $X_{h,1}$ or $%
X_{h,3}$, so that also $x_{h,2}=1$ (intersection between first or third row
and second column). This justifies the second equation. In general, we
require that the sums $\sum_{j}X_{i,j}$, with $j$\ running over the labels
with the same value of the same Boolean variable, are conserved across
different triplets (e. g. we understand that $x_{i,1}=x_{h,2}$ and $x_{h,2}$
are the same Boolean variable in triplets $i$\ and $h$). These linear
equations (representing the wires) are implemented by systems of gears
between the parts involved.

At this point the thought machine is completed. By applying a force to the
input part $Q$, the machine's motion from $Q=0~$to$~Q>0$ produces a solution
(provided that there is one, otherwise the machine is jammed) under the
simultaneous influence of all the problem constraints: in each triplet,
there is only one part that moves, the labels of all the parts that move
make up a Boolean assignment that solves the network.

\textbf{Acknowledgment }\textit{Thanks are due to Artur Ekert, David
Finkelstein, and Shlomit Finkelstein for encouragement and stimulating
discussions.}

\textbf{References}

Abbagnano N., Storia della Filosofia (1958). U.T.E.T., Volume I, p. 87.

Benioff, P. (1982). Quantum mechanical Hamiltonian models of Turing machines.%
\textit{\ J. Stat. Phys}., 29, 515.

Bennett, C.H. (1973). Logical reversibility of computation.\textit{\ IBM J.
Res. Dev.} 6, 525.

Bennett, C.H. (1982). The Thermodynamics of Computation -- a Review. \textit{%
Int. J. Theor. Phys.} 21, pp. 905.

Castagnoli, G. (1997). Quantum Computation Based on Retarded and Advanced
Propagation. quant-ph/9706019.

Castagnoli, G., Monti, D., and Sergienko, A. (1999). Performing quantum
measurement in suitably entangled states originates the quantum computation
speed up. quant-ph/9908015.

Castagnoli, G., and Finkelstein, D. (2001). Theory of the quantum speed up. 
\textit{Proc. Roy. Soc. Lond}. A 457, 1799. quant-ph/0010081.

Castagnoli, G. (2007). Quantum problem solving as simultaneous computation.
arXiv:0710.1744.

Castagnoli, G. (2008). On a fundamental problem solving mechanism explaining
the wholeness of perception.Toward a Science of Consciousness 2008 - Tucson
Discussions And Debates, eds. Hameroff, S. R., Kaszniak, A. W., Scott, A.
C., Cambridge, MA: MIT Press, p. 456.

Castagnoli, G. (2008). The mechanism of quantum computation. \textit{Int. J.
Theor. Phys., }vol. 47, number 8/August, 2008, 2181-2194.

Cleve, R., Ekert, A., Macchiavello, C., and Mosca, M. (1998). Quantum
Algorithms Revisited. Proc. Roy. Soc. Lond. A, vol. 454, number 1969,
339-354.

quant-ph/9708016.

Deutsch, D. (1985). Quantum theory, the Church-Turing principle and the
universal quantum computer. \textit{Proc. Roy. Soc}. (Lond.) A, 400, 97.

Faye, B., Laflamme, R., Mosca, M. (2007). An Introduction to Quantum
Computing. Oxford University Press.

Finkelstein, D. R. (1969). Space-Time Structure in High Energy Interactions.
Coral Gables Conference on Fundamental Interactions at High Energy. Center
of Theoretical Studies January 22-24, 1969. University of Miami. Timm
Gudehus, Geoffrey Kaiser, and Arnold Perlmutter Eds. Gordon and Breach,
Science Publishers, New York London Paris. pp. 324-343.

Finkelstein, D. R. (1969). Space-time code. \textit{Phys. Rev.} 184, 1261.

Finkelstein, D. R. (2008). Generational Quantum Theory. Preprint, to become
a Springer book.

Flew, A. (1984). A Dictionary of Philosophy. St. Martin's Press, New York.

Fredkin, E. and Toffoli, T. (1982). Conservative logic.\textit{\ Int. J.
Theor. Phys.} 21, 219.

Grover, L. K. (1996). A fast quantum mechanical algorithm for database
search. Proc. 28th Ann. ACM Symp. Theory of Computing.

Grover, L. K. (2001). From Schrodinger Equation to Quantum Search Algorithm.
Quant-ph/0109116.

Hameroff, S. R. and Penrose, R. (1996). Orchestrated Reduction Of Quantum
Coherence In Brain Microtubules: A Model For Cosciousness? Toward a Science
of Consciousness - The First Tucson Discussions And Debates, eds. Hameroff,
S. R., Kaszniak, A. W., Scott, A. C., Cambridge, MA: MIT Press, pp. 507-540.

Kaye, P., Laflamme, R., and Mosca, M. (2007). An Introduction to Quantum
Computing. Oxford University Press Inc., New York.

Mosca, M. and Ekert, A. (1999). The Hidden Subgroup Problem and Eigenvalue
Estimation on a Quantum Computer. \textit{Lecture Notes in Computer Science}%
, Volume 1509.

Mulligan, K. and Smith, B. (1988). Mach and Ehrenfels: The Foundations of
Gestalt Theory. From Barry Smith (ed.), Foundations of Gestalt Theory,
Munich and Vienna: Philosophia, 124.
http://ontology.buffalo.edu/smith/articles/mach/mach.html - N\_1\_

Penrose, R., (1994). Shadows of the Mind -- a Search for the Missing Science
of Consciousness. Oxford University Press.

Sheehan, D. (2008). Consciousness and the Physics of Time. Consciousness
Research Abstracts, proceedings of the Toward a Science of Consciousness
meeting, April 8-12, 2008, Tucson, Arizona.

Simon, D. (1994). On the Power of Quantum Computation. \textit{Proc. 35th
Ann. Symp. on Foundations of Comp. Sci.} 116--123.

Stapp, H. P. (1993). Mind, Matter, and Quantum Mechanics. Springer-Verlag,
Heidelberg, Berlin, New York.

\end{document}